\documentclass[runningheads]{svmult}

\usepackage{makeidx}   % allows index generation
\usepackage{graphicx}  % standard LaTeX graphics tool
\usepackage{subeqnar}  % subnumbers individual equations
\usepackage{multicol}  % used for the two-column index
\usepackage{physprbb}  % modified textarea for proceedings,
\makeindex             % used for the subject index
\begin{document}
\title*{Prospects for SNIa Explosion Mechanism\protect\newline Identification
Through SNRs}
\toctitle{Prospects for SNIa Explosion Mechanism\protect\newline Identification
Through SNRs}
\titlerunning{Clues to SNIa from SNRs}
\author{Carles Badenes\inst{1,2}
\and Eduardo Bravo\inst{1,2}
}
\authorrunning{Carles Badenes \& Eduardo Bravo}
\institute{
	Dpt. F\'\i sica i Eng. Nuclear, 
	Universitat Polit\`ecnica de Catalunya,
	Diagonal 647, \\
	08028 Barcelona, Spain
\and 
	Institut d'Estudis Espacials de Catalunya,
	Gran Capit\`a 2-4, 
	08034 Barcelona
}

\maketitle              % typesets the title of the contribution

\begin{abstract}
We present the first results from an ongoing work aimed to use supernovae
remnants to discriminate among different type Ia supernovae explosion models.
We have computed the hydrodynamic interaction of supernova ejecta with the 
interstellar medium, obtaining the evolution of the density, temperature and ionization 
structure of the remnant. We have used ejecta profiles obtained from 1D hydrodynamic 
calculations of the different explosion mechanisms that are currently under debate.
We have analyzed the best indicators that allow to discriminate
among the different explosion mechanisms, taking into account the diversity of scenarios 
proposed for the presupernova evolution of the binary system, and the uncertain amount of 
electron heating in collisionless shocks.
\end{abstract}

\section{Thermonuclear supernovae remnants}
Modern X-ray
observatories, like {\sl Chandra} and {\sl XMM-Newton} give the opportunity to
collect a huge amount of high quality, high resolution, data about supernova
remnants (SNRs). \index{supernova remnants}
Whereas the atomic
codes that allow to compute and fit the X-ray spectra have experienced a
considerable improvement in the last years, this is not the case of our
knowledge of the evolution of SNRs. 
One step towards a more complete understanding of these objects is to couple
one-dimensional hydrodynamical calculations to an accurate ionization code.

The explosion mechanism(s) responsible for thermonuclear supernovae (SNIa) is still a
matter of strong debate. In spite of the advances made in recent years
in the understanding of the physics of the flame, there is still no definitive
answer to the question of what is the mode of propagation of a thermonuclear
flame through a massive C+O white dwarf. To answer this, we need to exploit every piece of 
evidence that can hold a signature of the flame propagation history. Here, we 
present the first results from an ongoing work aimed to use SNRs to discriminate among different SNIa explosion models. We have made 
a systematic exploration of the parameter space of  
explosion mechanisms including pure detonations, pure deflagrations, delayed 
detonations, pulsating delayed detonations, and sub-Chandrasekhar models, and we
have obtained a set of explosion ejecta profiles with the aid of a 1D SN
hydrodynamics code. Then, for each one of the profiles, we have computed the 
hydrodynamic interaction of the supernova ejecta with a uniform interstellar 
medium (ISM), obtaining the evolution of the density, temperature and ionization
structure of the remnant. In order to be sure that SNR properties allow to discriminate between
explosion models it is necessary to look for the effects of other unknowns of the problem, 
mainly the presupernova evolution, which determines the ambient medium structure at the time
of SN explosion, and the physics of collisionless shocks.

\subsection{Presupernova Evolution}
\begin{figure}
\begin{center}
{
 \centering 
 \leavevmode 
 \includegraphics[width=.3\textwidth, angle=90]{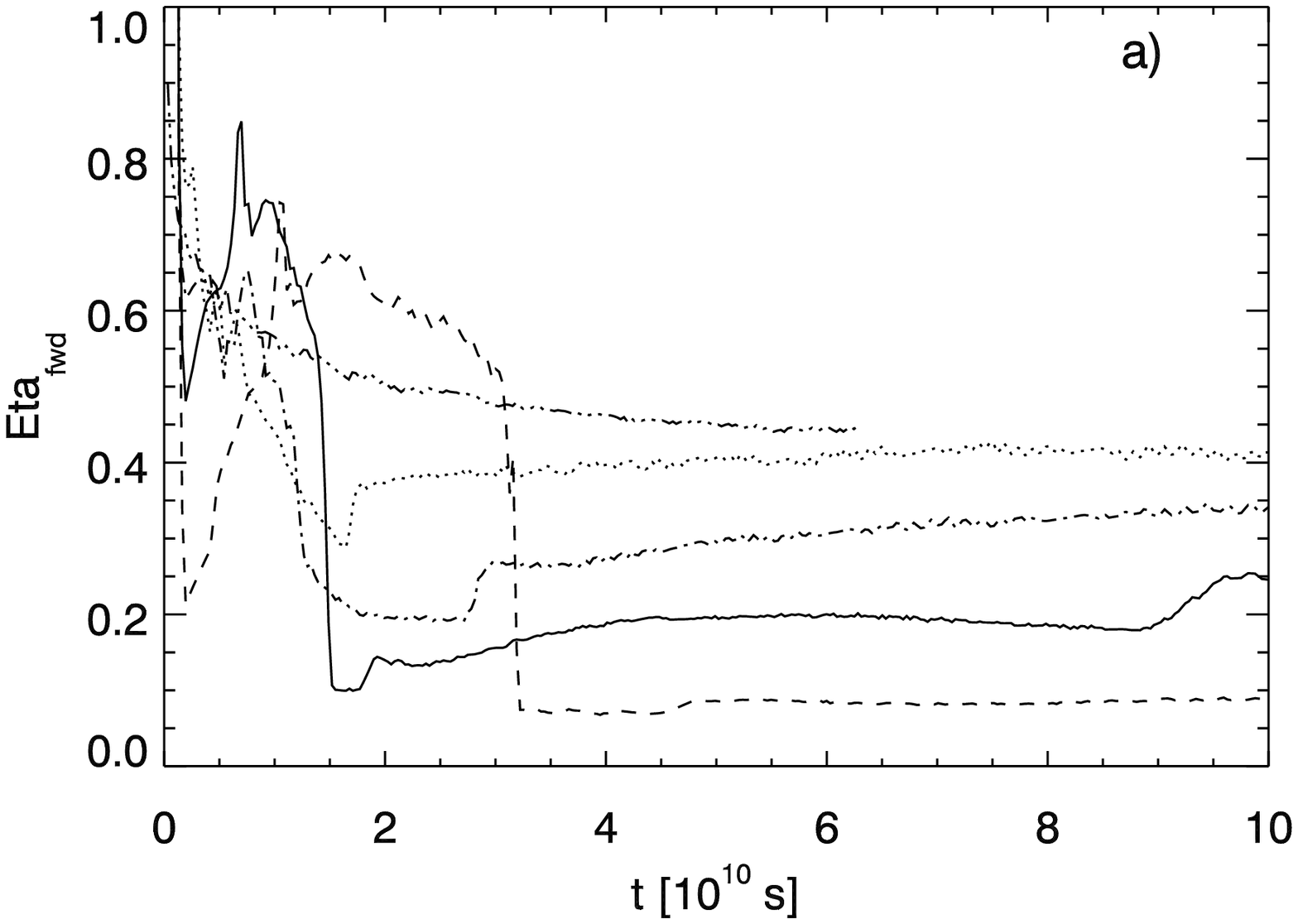}
 \hfil 
 \includegraphics[width=.3\textwidth, angle=90]{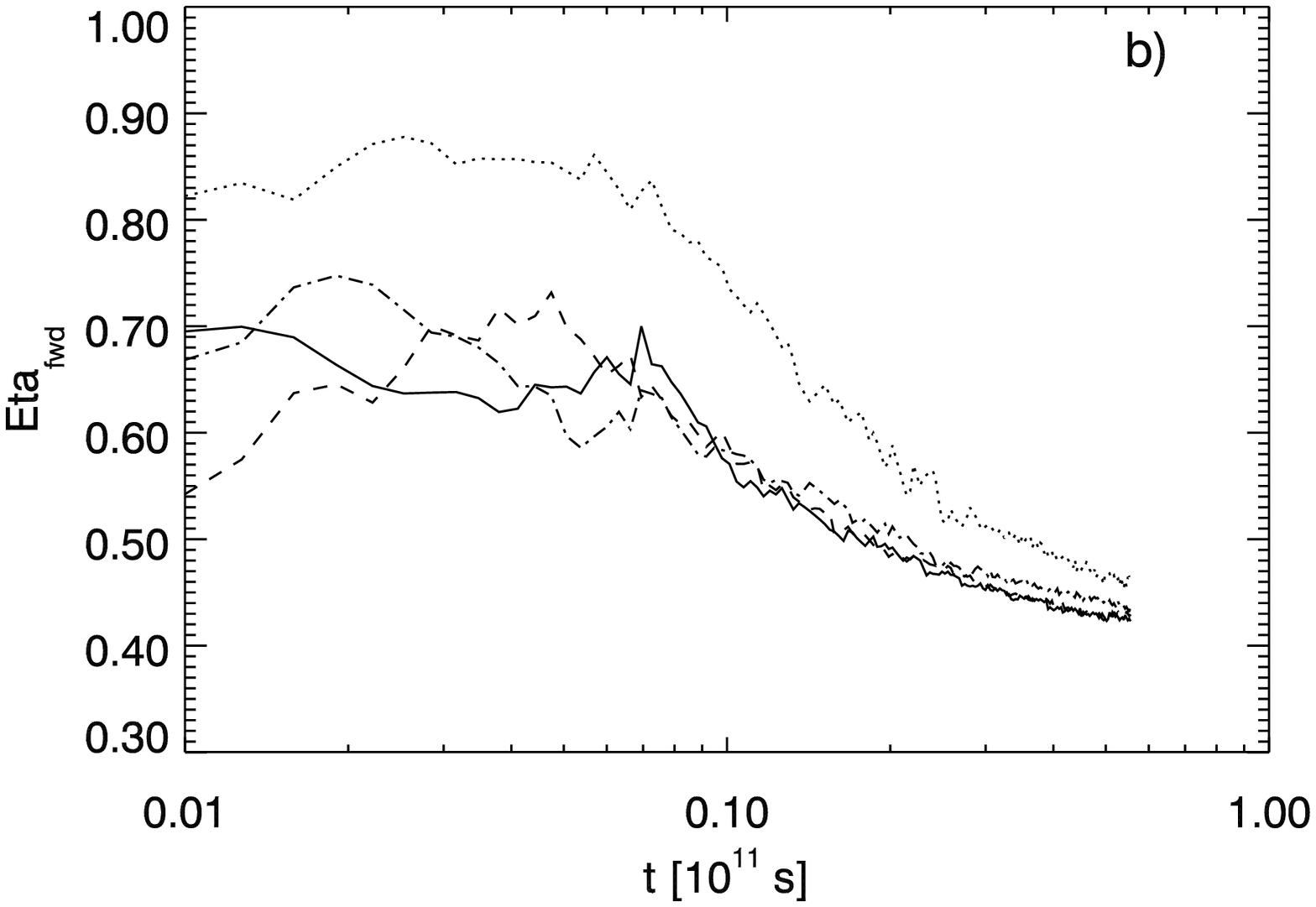}
}
\end{center}
\caption{Expansion parameter of the forward shock as a function of time. 
(\textbf{a}) Wind models: A ({\it solid line}), B ({\it dotted line}), C ({\it dashed
line}), and D ({\it dash-dotted line}), and uniform ISM 
({\it triple-dot-dashed line}). 
(\textbf{b}) Explosion models: Delayed detonation \index{delayed detonation} ({\it solid line}), deflagration ({\it dotted line}), 
pulsating delayed detonation ({\it dashed line}), and sub-Chandrasekhar 
({\it dash-dotted line})
}
\label{fig1}
\end{figure}

We have tested four simple models of presupernova \index{presupernova} mass loss 
due to binary wind \index{wind} (for details see 
\cite{bb01} and \cite{bb02}), with parameters chosen in order to reproduce the
gross features of different evolutionary scenarios 
(references can be found in \cite{bb01}).
The structure of the ambient medium at the time of SN explosion has been computed with a SNR
hydrocode which included radiative losses. 
Presupernova winds interact with the ISM 
producing the typical forward--shock/contact--discontinuity/reverse--shock structure, 
and providing:\begin{itemize}
\item
A cold dense shell appropriated for the formation of dust. This shell could also
play a role in the formation of a light echo.\index{light echo}
\item
A close circumstellar shell \index{circumstellar shell} which could interact early with the SN ejecta (but
only in our wind model 
D: low velocity wind active till the SN explodes). 
\end{itemize}
We have computed the interaction of SN ejecta with the ambient medium sculpt\-ed by the four wind 
models, together with the interaction of the same ejecta profile with a constant density ISM.  
In Fig.~\ref{fig1} we show the expansion parameter \index{expansion parameter} of the forward shock 
(defined as 
$\eta=\D(\ln R_{\mathrm shock})/\D(\ln t)$). 
As can be seen, the dynamical properties of SNR are mostly sensitive to the presupernova 
history at ages larger than that of Tycho''s SNR ($1.357\times10^{10}$~s).

\subsection{Explosion Mechanism}
We have explored different explosion mechanisms, \index{explosion mechanism} whose ejecta are characterized 
by their density and chemical profiles, obtained with a SNIa 1D hydrocode
(see \cite{bb02} for details).
We have computed the interaction of each SN ejecta profile with a uniform
density ISM ($\rho_{\mathrm ISM} = 10^{-24}$~g$\cdot$cm$^{-3}$) with the 
same SNR hydrocode previously described, and we have obtained the ionization
structure of the SNR as a function of time.
In principle, there are two ways to discriminate between the different ejecta 
profiles: one is the dynamic evolution of the SNR, the other is its X-ray emission. 

\subsubsection{Dynamical evolution of the supernova remnant}
The dynamics of young SNRs is sensitive to the explosion mechanism up to a 
few hundred years after the explosion. As can be seen in Fig.~\ref{fig1}, our
main result is that
the expansion parameter of the deflagration \index{deflagration} model is the largest at early times due to the sharp density contrast at the point of flame quenching.

\subsubsection{Collisionless Shocks Physics}
\begin{figure}[b]
\begin{center}
{
 \centering 
 \leavevmode 
 \columnwidth=.45\columnwidth 
 \includegraphics[width=.3\textwidth, angle=90]{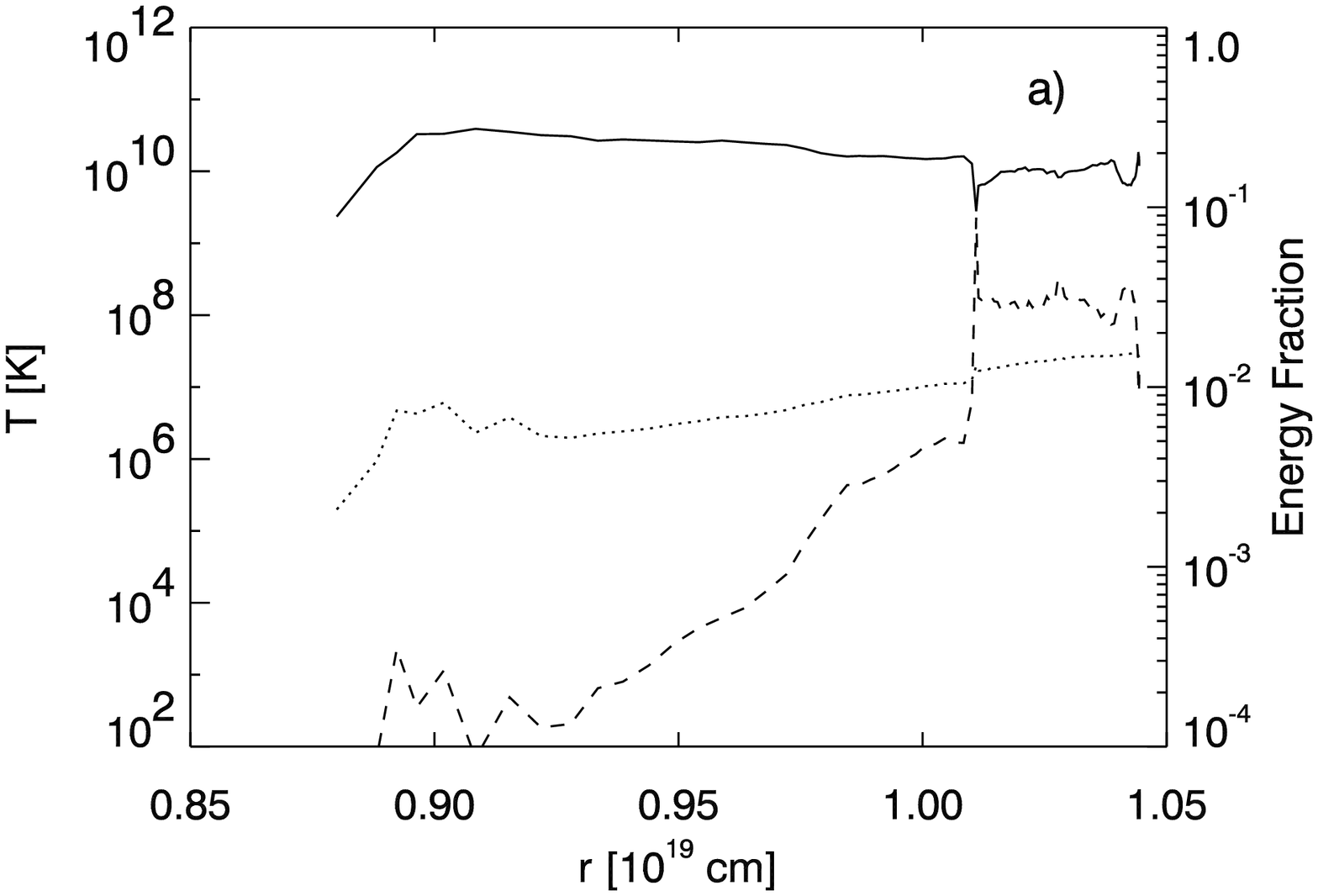}
 \hfil 
 \includegraphics[width=.3\textwidth, angle=90]{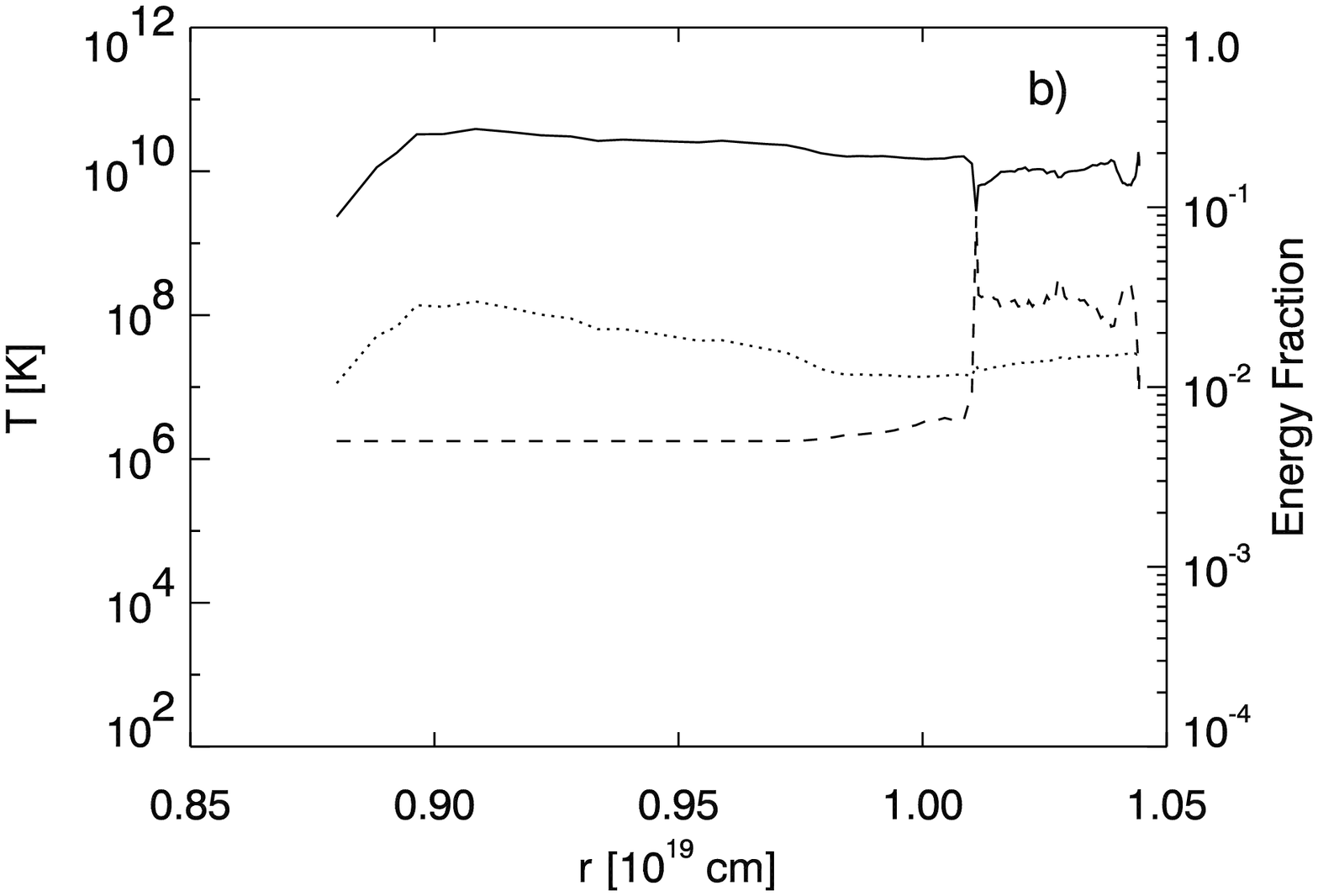}
}
\end{center}
\caption{Ion ({\it solid line}) and electron ({\it dotted line}) temperatures in the region between the reverse shock
(left) and the contact discontinuity (right), together with the fraction of
internal energy in the electron gas ({\it dashed line}) for the delayed
detonation model at the age of Tycho's SNR. (\textbf{a}) Assuming
no electron heating at the shock front. (\textbf{b}) Assuming a moderate amount
of electron heating at the shock front (see text for details)}
\label{fig2}
\end{figure}

\index{collisionless shocks}
The physical processes that are at work in 
collisionless shocks physics are not well understood, the main uncertainty being the amount of
electron heating in the shocks. The heavy 
plasma ejected by SNIa poses additional problems, because the electron number density
depends on the local chemical composition and the degree of ionization
\index{ionization} of each
element (which is out of ionization equilibrium, \cite{blr01}). 

We have computed carefully the ionization evolution of the plasma, and have
checked the
sensitivity of the results to the amount of electron heating at the shock. 
In Fig.~\ref{fig2} we 
show the results from two calculations in which we have assumed either no electron heating 
at the shock or a moderate electron heating (1\% of the ion temperature).
Coulomb interactions between electrons and ions were included in both
calculations.
Our main results concerning the influence of electron heating at the shock are:
\begin{itemize}
\item
The ionization timescale does not depend on it at all.
\item
The electron temperature \index{electron temperature} near the reverse shock (and, thus, the high energy X-ray \index{X-rays}
continuum) depends strongly on it.
\end{itemize}

\subsubsection{X-ray emissivity}
\begin{figure}[t]
\begin{center}
{
 \centering 
 \leavevmode 
 \includegraphics[width=.2\textwidth, angle=90]{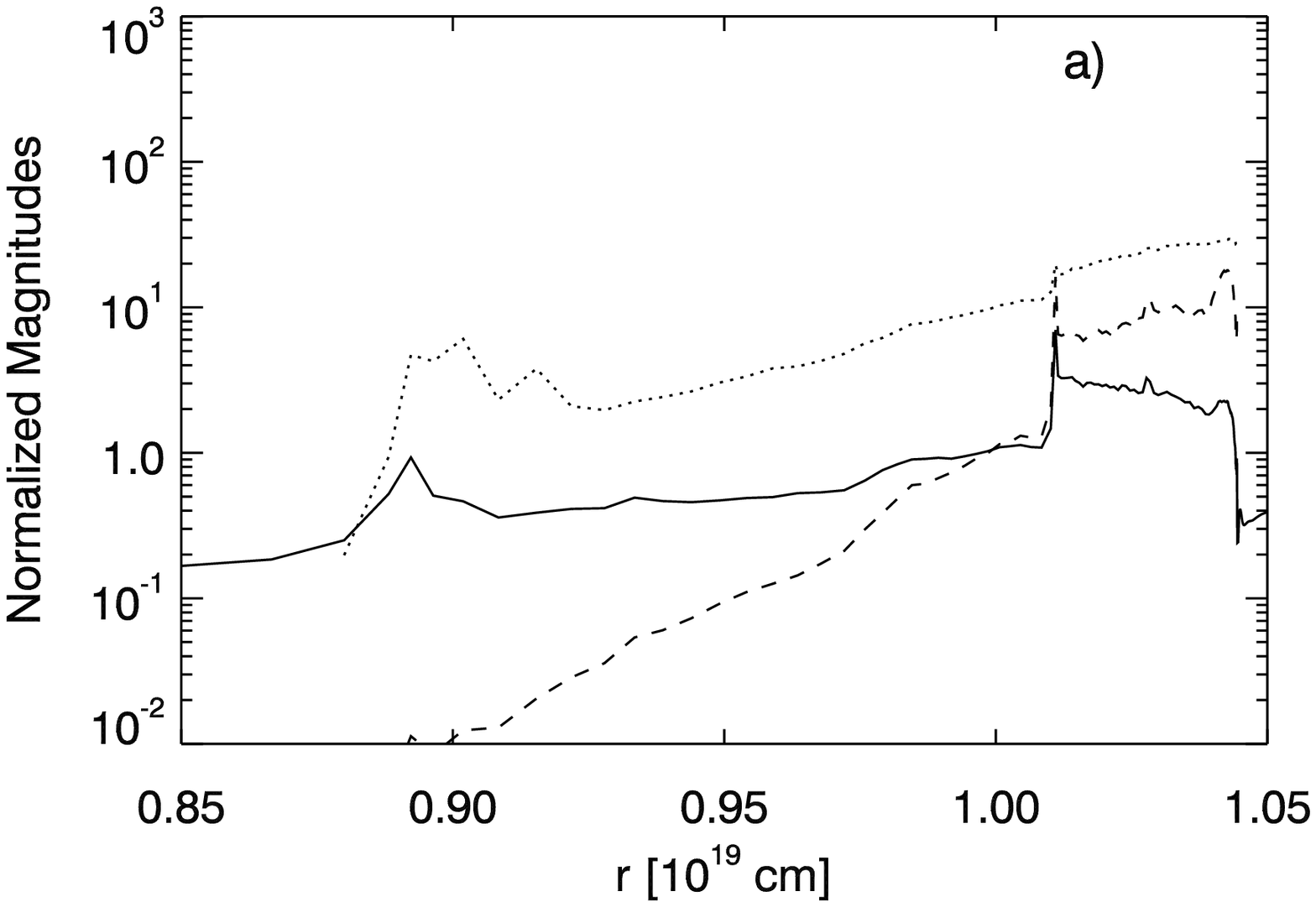}
 \hfil 
 \includegraphics[width=.2\textwidth, angle=90]{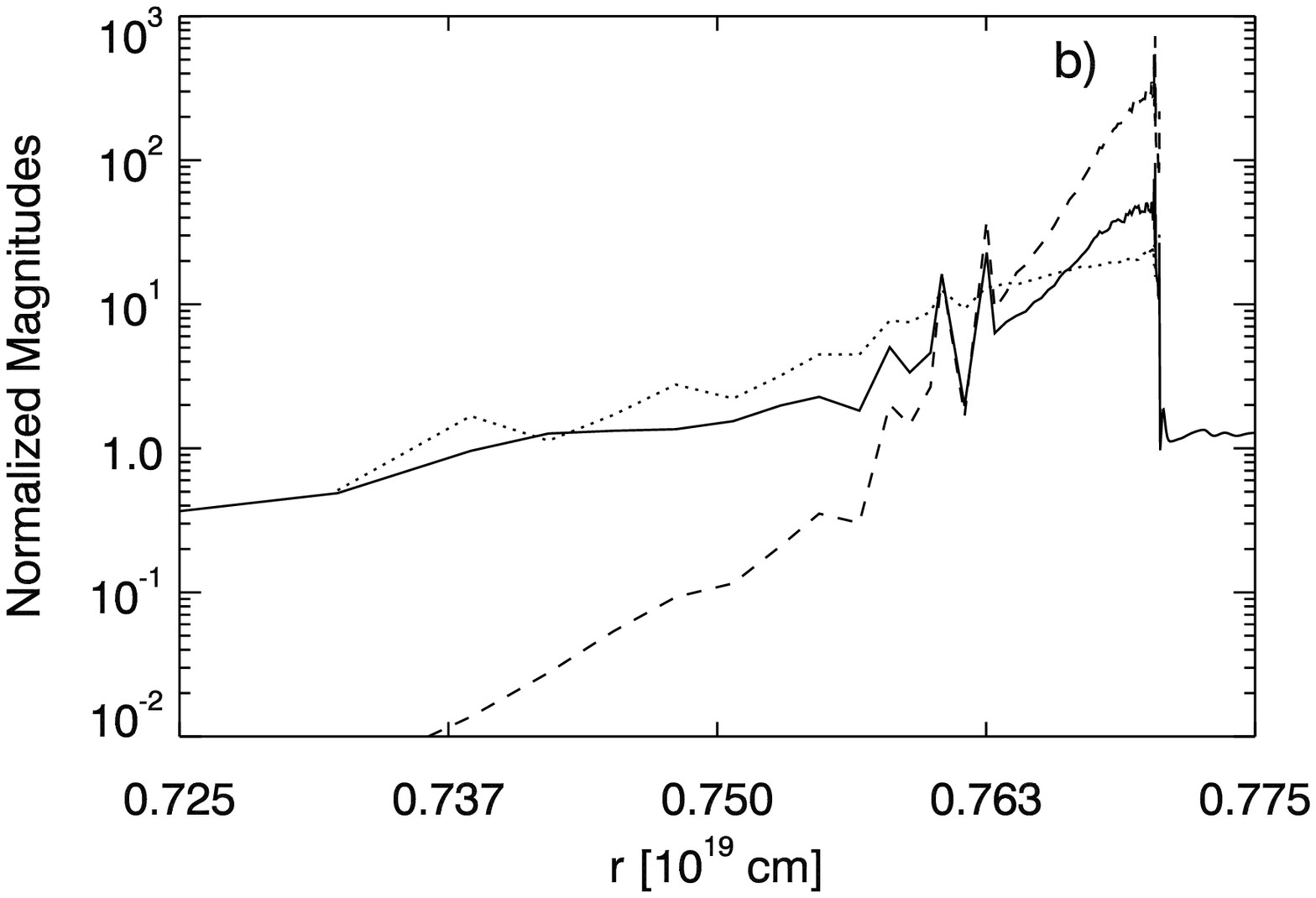}
 \hfil 
 \includegraphics[width=.2\textwidth, angle=90]{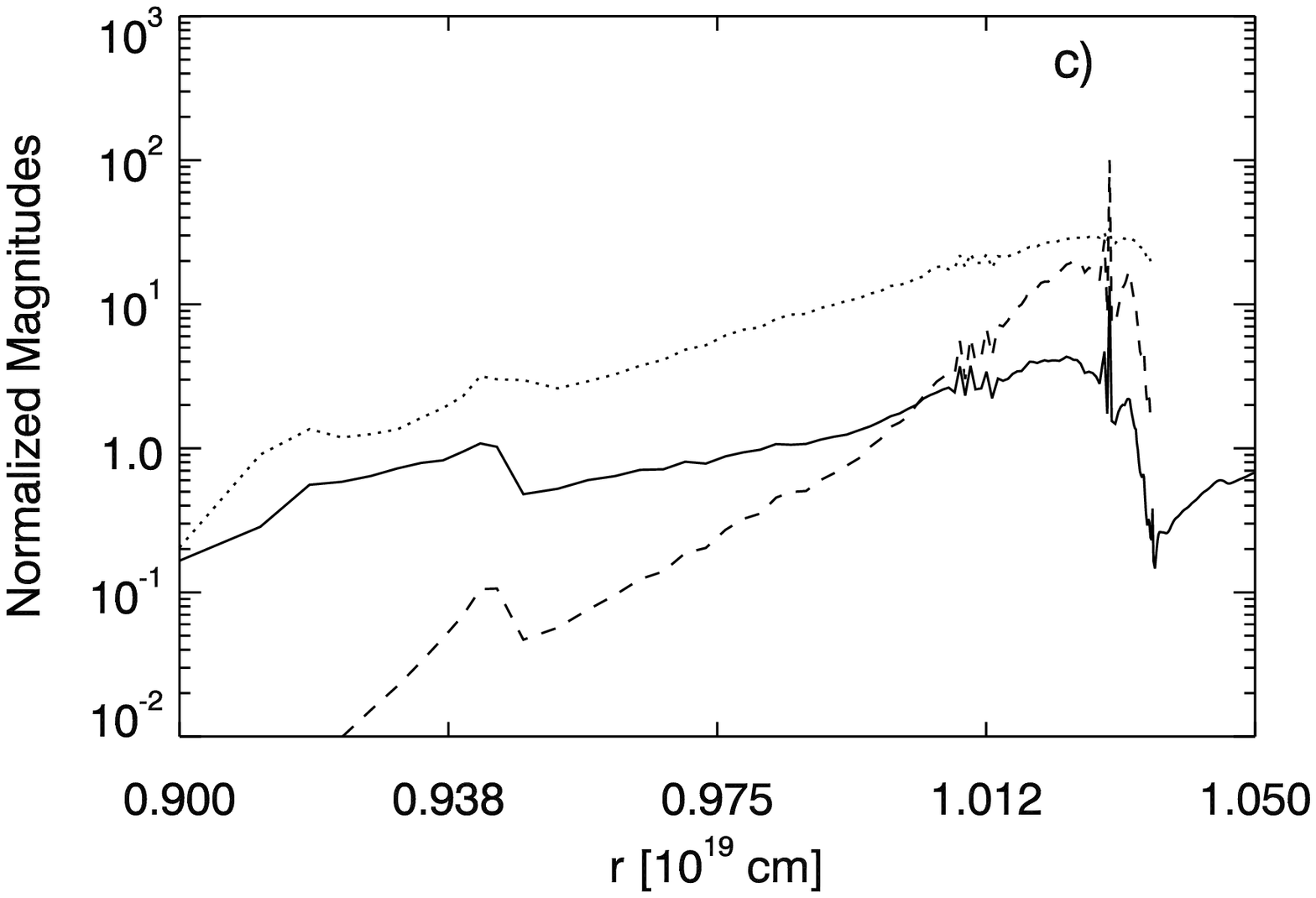}
}
\end{center}
\caption{
Normalized density ($\rho/10^{-24}$~g$\cdot$cm$^{-3}$, {\it solid line}), 
electron temperature ($T_{\mathrm e}/10^6$~K, {\it dotted line}), and
ionization timescale ($n_{\mathrm e}t/10^9$s$\cdot$cm$^{-3}$, {\it dashed
line}), 
in the region between the
reverse shock (left) and the contact discontinuity (right). Note the different $r$ scales. 
(\textbf{a}) Delayed detonation model. (\textbf{b}) Deflagration model. 
(\textbf{c}) Pulsating delayed detonation model
}
\label{fig3}
\end{figure}

Finally, we address the problem of the X-ray emissivity associated to each explosion model. 
Up to now, we have not concluded this part of the project, but some preliminary results 
are already available.
In Fig.~\ref{fig3} we show the density, temperature and ionization timescale profiles 
of the deflagration, pulsating delayed detonation and delayed detonation models at the age of Tycho's
SNR. Our results
show that the most model sensitive characteristics are:
\begin{itemize}
\item
The ionization timescale close to the contact discontinuity and, thus, the centroid of
the Fe/S He$\alpha$ complex of X-ray lines.
\item
The density profile and, thus, the emission measure and the X-ray luminosity.
\item
The high energy continuum is mainly dependent on the assumed physics of the
collisionless shocks.
\end{itemize}

\noindent {\small This work has been supported by the MCYT grants EPS98--1348 and AYA2000--1785 
and by the DGES grant PB98-1183--C03-02. C. B. is very indebted to  
CIRIT for a grant.}

%INDEX%%%%%%%%%%%%%%%%%%%%%%%%%%%%%%%%%%%%%%%%%%%%%%%%%%%%%%%%%%%%%%%
% Please check with the editor of your book whether he plans to
% include a "mutual" subject index - if so, please code your entries
% in the standard syntax. For your own purposes you may print your
% "personal" index by using the following commands:
%
%\clearpage
%\addcontentsline{toc}{section}{Index}
%\flushbottom
%\printindex
%%%%%%%%%%%%%%%%%%%%%%%%%%%%%%%%%%%%%%%%%%%%%%%%%%%%%%%%%%%%%%%%%%%%%

\end{document}